\documentclass[prb,twocolumn,showpacs]{revtex4-1}
\usepackage[dvips]{graphicx}
\usepackage{epsfig}
\usepackage{color}
\usepackage[normalem]{ulem}
\usepackage{amsmath,amsfonts,amssymb,bm}
\setcitestyle{numbers,square}

\begin{document}
\title{Comment on "Many-body localization in Ising models with random long-range interactions"}
\author{Andrii O. Maksymov, Noah Rahman, Eliot Kapit,  Alexander L. Burin}
\affiliation{Tulane University, New
Orleans, LA 70118, USA}

\date{\today}
\begin{abstract}
This comment is dedicated to the investigation of many-body localization  in a quantum Ising model with long-range power law interactions, $r^{-\alpha}$, relevant for a variety of systems ranging from electrons in Anderson insulators to spin excitations in chains of cold atoms. It has been earlier argued \cite{ab06preprint,ab15MBL}  that this model obeys the dimensional constraint suggesting the delocalization of all finite temperature states in  thermodynamic limit for $\alpha \leq 2d$ in a $d$-dimensional system. This expectation conflicts with the recent numerical studies of the specific interacting spin model in Ref.  \cite{Li16TransvIs}. To resolve this controversy we reexamine the model  of Ref. \cite{Li16TransvIs} and demonstrate that the infinite temperature states there obey the dimensional constraint. The earlier developed scaling theory for the critical system size required for delocalization \cite{ab15MBL} is extended to small exponents $0 \leq \alpha \leq d$.  Disagreements between two works  are explained by the non-standard selection of investigated states in the ordered phase and misinterpretation of the localization-delocalization transition in Ref. \cite{Li16TransvIs}. 
\end{abstract}

\maketitle

\section{Introduction}
\label{sec:intro}

The many-body localization (MBL) transition separates two distinguishable thermodynamic behaviors. The delocalized system acts as a thermal bath for each small part of it  \cite{Huse15Thermalization,Cohen13} while in the localized system its different parts are approximately independent and can be characterized by related local integrals of motion \cite{Huse14IntMot}. Recent experimental investigations of many-body localization \cite{Monro16,LukinDiamond16} are carried out in systems of interacting spins coupled by the long-range interaction decreasing with distance according to power law $U(r) \propto r^{-\alpha}$. The interaction exponent $\alpha$ can be modified experimentally \cite{Lukin14MBLGen,Yao14MBLLongRange} and it is important to understand the effect of power-law interactions on localization. 

According to the previous work of one of the authors (with coworkers) \cite{ab06preprint,ab98book} (see also more recent work, Refs.   \cite{Yao14MBLLongRange,ab16GutmanMirlin,ab15MBL,ab15MBLXY}) the MBL problem in  systems with power law interactions is subject to a dimensional constraint.  This constraint suggests that localization is not possible in the thermodynamic limit of an infinite system at a finite temperature if the interaction decreases with the distance slower than $1/r^{2d}$ (where $d$ is a system's dimension in the case of mixed Ising-Heisenberg interactions). 
Dimensional constraints have been derived assuming that $\alpha \geq d$ to avoid the single-particle delocalization. 

However, recently a disagreement with this dimensional constraint has been reported in Ref.  \cite{Li16TransvIs} for the quantum Ising model with the long-range interactions for $\alpha=0.5$, $1$ and $1.5$. In this paper the spin chain of $N$ spins described by the Hamiltonian 
\begin{eqnarray}
\widehat{H}=J\sum_{1\leq i< j\leq L}\frac{(1+h_{i}h_{j})}{|j-i|^{\alpha}}\sigma_{i}^{z}\sigma_{j}^{z}+B\sum_{i=1}^{L}\sigma_{i}^{x}
\label{eq:H}
\end{eqnarray}
has been investigated, where spins are represented by Pauli matrices $\sigma$ and disorder is introduced using random  parameters $h_{i}$ uncorrelated in different sites $i$ and uniformly distributed within the domain $(-W, W)$ while the parameter $W$ describes the effective disorder. In all studies the transverse field, $B$, has been set to  $0.6J$. 

The localization transition has been investigated in Ref. \cite{Li16TransvIs} using level statistics and entanglement entropy. The level statistics have been characterized using the averaged ratio of successive gaps, $<r>$, defined as \cite{OganesyanHuse07} 
\begin{eqnarray}
<r> = \left<  \frac{{\rm min} (\delta_{n}, \delta_{n+1})}{{\rm max}(\delta_{n}, \delta_{n+1})} \right>,
\label{eq:Og}
\end{eqnarray}
where $\delta_{n} =E_{n+1}-E_{n}$ is the energy difference of adjacent energy levels of the system, Eq. (\ref{eq:H}), obtained by means of exact diagonalization. According to Ref. \cite{OganesyanHuse07} in the delocalized regime characterized by Wigner-Dyson statistics  one has $<r>\approx 0.5307$ while in the case of localization where the Poisson statistics is expected one has $<r> \approx 0.3863$.  The consideration has been limited to the eigenstates with energies close to the middle energy between the minimum and maximum energies $E_{\rm middle}=(E_{\rm min}+E_{\rm max})/2$ represented by the dimensionless parameter  (see Ref. \cite{Li16TransvIs} for detail)
\begin{eqnarray}
\epsilon=\frac{E-E_{\rm min}}{E_{\rm max}-E_{\rm min}}\approx \frac{59}{120}.
\label{eq:Og1}
\end{eqnarray}

Based on the analysis of the level statistics parameter $<r>$ at $\epsilon \approx 59/120$ the authors found the localization at any disordering $W$ for the smallest interaction exponents $\alpha = 1/2$, $1$ and size-independent localization-delocalization transition for $\alpha=1.5$ in contrast to the earlier suggested dimensional constraint \cite{ab06preprint}. The  results for $\alpha=0.5$ have been found consistent with the earlier work \cite{Hauke15MBLLongRange} where localization was observed in a different model of interacting spins with the power-law interaction $1/r^{0.5}$. 

Since systems with long-range interactions are of both acute fundamental and and experimental \cite{Monro16,LukinDiamond16} interest due to the ubiquity of charge, dipole, magnetic and elastic forces  \cite{ab98book,ab06preprint}, it is important to understand and interpret  the conflict between the qualitative analysis \cite{ab98book,ab06preprint,Yao14MBLLongRange,ab15MBL,ab16GutmanMirlin} leading to the aforementioned dimensional constraint and the numerical results of Ref. \cite{Li16TransvIs}. The consideration of the earlier work \cite{ab15MBL} has been limited to the interaction exponents $\alpha \geq d$ and its extension to smaller exponents $0\leq \alpha < d$ is another problem of interest. These problems are investigated in the present work. 

Below we show that the discrepancy between two approaches originates from the specifics of the model, Eq. (\ref{eq:H}), considered in Ref. \cite{Li16TransvIs} and the associated choice of the representative energy $E_{\rm middle}$. In the case of $\alpha \leq 1$ the maximum eigenstate energy, $E_{\rm max}$, for the spin chain increases superlinearly with the number of spins $N$.  Indeed, for the ferromagnetic state $\sigma_{i}^{z}=1$ or the alternative maximum at $\sigma_{i}^{z}={\rm sign} (h_{i})$ this energy scales as $E_{\rm max} \propto N^{2-\alpha}$ for $\alpha < 1$ or $N\ln(N)$ for $\alpha=1$, while the minimum energy scales as $|E_{\rm min}| \sim N$. Consequently, the energy of the eigenstates investigated in Ref. \cite{Li16TransvIs} scales as   $E\sim E_{\rm max}/2$.  In the case of small interaction exponents, $\alpha \leq 1$, the states with this energy belong to the ordered phase of the system possessing very small many-body density of states (see Fig. \ref{fig:DoS}), where delocalization is substantially suppressed (see Fig. \ref{fig:OgMiddleW1}, cf. Refs. \cite{Laumann14,ab16preprintSG}). We show that the dimensional constraint remains valid for the states with energies close to zero corresponding to infinite temperature even in the model of Ref. \cite{Li16TransvIs} in contrast to the states at energy $E \approx E_{\rm middle}$. 
 
The dimensional constraint for the states with energies close to zero (corresponding to infinite temperature) is investigated in Sec. \ref{sec:Zer} where the results of Ref. \cite{Hauke15MBLLongRange} are also considered briefly. 
The analysis of the middle energy states is performed  in Sec. \ref{sec:Middle}, whereupon we conclude. 


\section{Zero energy states ($T=\infty$)}
\label{sec:Zer}

In the infinite temperature limit of the model, Eq. (\ref{eq:H}) all spins are assumed to be non-correlated. Then the consideration of Ref. \cite{ab16preprintSG} can be applied to the present problem. Either strong or weak interaction regimes are applicable. Then the critical randomness parameter, $W$, should increase to infinity in the thermodynamic limit $N \rightarrow \infty$. Consequently in the limit of interest of large number of spins, $N$, one can neglect the unity term compared to the product $h_{i}h_{i}$ in the definition of the interaction, Eq. (\ref{eq:H}). 

Here we briefly repeat some qualitative arguments from Ref. \cite{ab16preprintSG}.   Each spin $i$ is subjected to a longitudinal field $\Phi_{i}=\sum_{j}J_{ij}\sigma_{j}^{z}$, where $J_{ij}=Jh_{i}h_{j}/|i-j|^{\alpha}$. For fully random spin projections (infinite temperature limit) the longitudinal field $\Phi_{i}$ is zero in average and it is distributed nearly uniformly within the domain $(-\sigma, \sigma)$ where the size of the domain can be estimated as 
\begin{eqnarray}
\sigma \sim \sqrt{\sum_{j}J_{ij}^2} \sim 
\begin{cases}
    JW^2, & \text{if $\alpha > d/2$},\\
    JW^2\sqrt{\ln(N)}, & \text{if $\alpha=d/2$},\\
    JW^2 N^{1/2-\alpha/d}, & \text{if $\alpha<d/2$}.\\
  \end{cases}.    
\label{eq:Rand}
\end{eqnarray}

The localization-delocalization transition is associated with resonant spins $i$ satisfying the condition $|\Phi_{i}| < B$ and the probability of such resonance can be estimated as $P_{res} \sim B/\sigma$. The total number of spin resonances per state is given by $N_{res} =NP_{res}$. According to Ref. \cite{ab16preprintSG} the delocalization transition can be determined by the condition $N_{res}\ln(J_{N}/B) \sim 1$ in the case of strong interaction, $J_{N} =J/N^{\alpha} > B$. It turns out that the interaction is indeed strong for $\alpha < d$ and the localization delocalization transition is determined as 
\begin{eqnarray}1 \sim
\begin{cases}
    N\frac{B}{JW_{c}^2}\ln(N^{1-\alpha/d}), & \text{if $d>\alpha > d/2$},\\
    N\frac{B}{JW_{c}^2}\sqrt{\ln(N)}, & \text{if $\alpha=d/2$},\\
    N^{1/2+\alpha/d}\frac{B}{JW_{c}^2}\ln(N), & \text{if $\alpha<d/2$},\\
  \end{cases}    
\label{eq:LocTrStr}
\end{eqnarray}
where the parameter $W_{c}$ estimates the critical randomness corresponding to the localization transition. 

\begin{figure}[h!]
\centering
\includegraphics[width=\columnwidth]{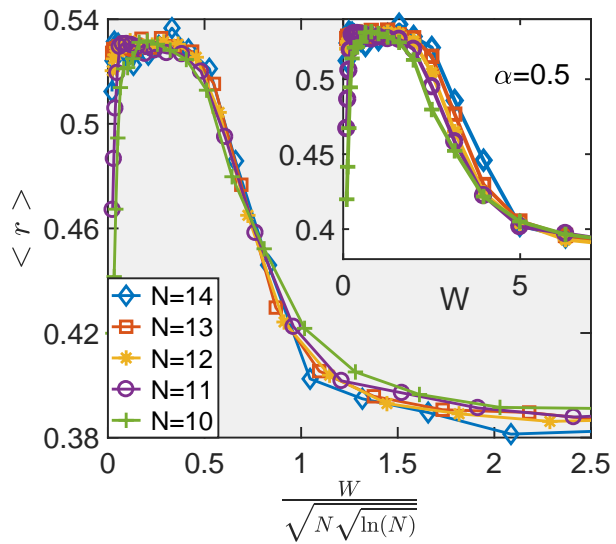}
\caption{\small The level statistics ($<r>$) vs. disordering  $W$ (inset) or rescaled disordering according to Eq. (\ref{eq:LocTrStr}) for interaction exponent $\alpha=0.5$ and different numbers of spins $N=10$, $11$, $12$, $13$ and $14$.}
\label{fig:0_5mid}
\end{figure}

The case of $\alpha \geq d$ corresponding to the weak interaction regime has been considered in Ref. \cite{ab16preprintSG}. In that case it has been found in accordance with earlier studies \cite{ab06preprint} that the delocalization inevitably takes place in the thermodynamic limit for $\alpha <2d$ and the critical disordering, $W_{c}$  at a finite number of spins $N$ is determined as 
\begin{eqnarray}
N^{2-\alpha/d}B \sim JW_{c}^2.    
\label{eq:LocTrWk}
\end{eqnarray}
Both estimates in Eqs. (\ref{eq:LocTrStr}) and (\ref{eq:LocTrWk}) are valid assuming $W_{c} \gg 1$ which takes place at a sufficiently large number of spins ($N$).

\begin{figure}[h!]
\centering
\includegraphics[width=\columnwidth]{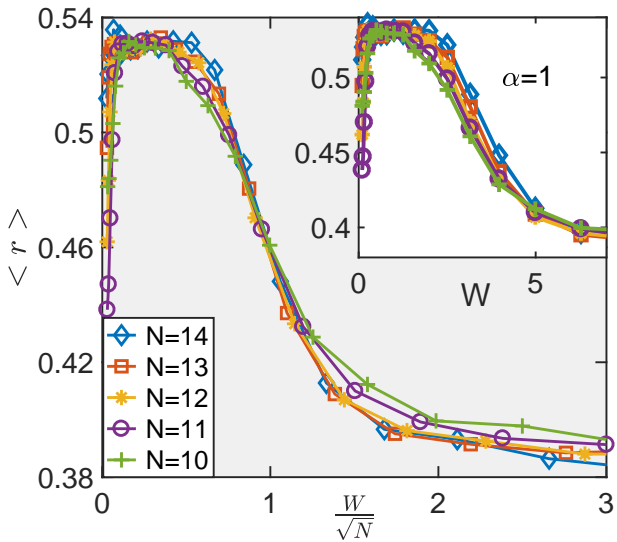}
\caption{\small The level statistics ($<r>$) vs. disordering   $W$  (inset) and rescaled disordering for $\alpha=1$.}
\label{fig:1mid}
\end{figure}

\begin{figure}[h!]
\centering
\includegraphics[width=\columnwidth]{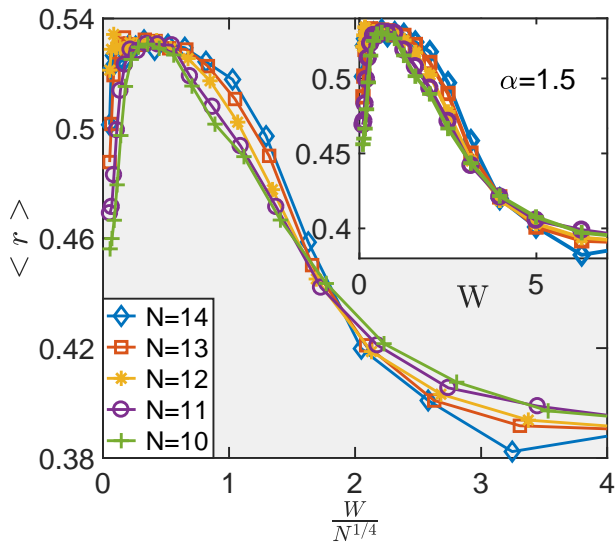}
\caption{\small The level statistics ($<r>$) vs. disordering   $W$  (inset) and rescaled disordering for $\alpha=1.5$.}
\label{fig:1_5mid}
\end{figure}

To verify the theoretical  predictions, Eqs. (\ref{eq:LocTrStr}) and (\ref{eq:LocTrWk}),  we analyzed numerically the level statistics performing exact diagonalization  of the Hamiltonian, Eq. (\ref{eq:H}), for the states at zero energy, corresponding to the infinite temperature limit and for numbers of spins $10 \leq N \leq 14$. The results are presented in Figs. \ref{fig:0_5mid}, \ref{fig:1mid} and \ref{fig:1_5mid}  for power law interaction exponents $\alpha=0.5$, $1$ and $1.5$, respectively, as in Ref. \cite{Li16TransvIs}. All results are given for the states of even parity with respect to the symmetry transformation $\sigma^{z} \rightarrow -\sigma^{z}$ of the Hamiltonian $1$. The results for the odd parity are  quite similar. All curves are averaged over $1000$ realizations of random interactions as in Ref. \cite{Li16TransvIs}. 

 In all three cases the delocalization clearly takes place at sufficiently small disordering $W$ where the average ratio parameter has a plateau at $<r> \approx 0.53$. This contrasts with the statements of Ref. \cite{Li16TransvIs} for small interaction exponents $\alpha=0.5$ and $1$ and the reason for this discrepancy is the difference in the energy of the considered system states as detailed below in Sec. \ref{sec:Middle}. At very small disordering, $W \leq 0.1$, the ratio parameter  deviates from the plateau. This is due to the system reflection symmetry at $W=0$, which breaks Hamiltonian into two non-interacting blocks having no level repulsions. There is no localization for $W \rightarrow 0$ as was verified by the analysis of participation ratios of eigenstates, which is comparable to the total number of states and does not even decrease at $W\rightarrow 0$ (not shown here). 

In all three cases it is expected that the localization threshold $W_{c}$ approaches infinity in the thermodynamic limit ($N\rightarrow \infty$) as $W_{c} \sim \sqrt{N\ln(N)}$ for $\alpha =0.5$, $W_{c} \sim \sqrt{N}$ for $\alpha=1$ and $W\sim N^{1/4}$ for $\alpha =1.5$ (see Eqs. (\ref{eq:LocTrStr}) and (\ref{eq:LocTrWk})). Indeed, the shifts of the transition between delocalization ($<r>\approx 0.53$) and localization ($<r> \approx 0.38$) regimes towards  larger disordering $W$ is seen in all three cases (insets in Figs. \ref{fig:0_5mid},  \ref{fig:1mid} and \ref{fig:1_5mid}). To examine the relevance of the transition point dependence on size (Eqs. (\ref{eq:LocTrStr}) and (\ref{eq:LocTrWk})) we rescaled a disordering as indicated in $x$-axes in Figs. \ref{fig:0_5mid}, \ref{fig:1mid} and \ref{fig:1_5mid}  similarly to Ref. \cite{ab15MBL}. For small interaction exponents, $\alpha=0.5$ and $1$, this rescaling places the transitions to nearly the same curve confirming the qualitative predictions of Eqs. (\ref{eq:LocTrStr}) and (\ref{eq:LocTrWk}). The situation is less conclusive for $\alpha=1.5$ possibly because the dependence $W_{c} \sim N^{1/4}$ is weak and the finite size effects are significant there for $N \leq 14$ (cf. Refs. \cite{ab15MBL,ab15MBLXY}).

\begin{figure}[h!]
\centering
\includegraphics[width=\columnwidth]{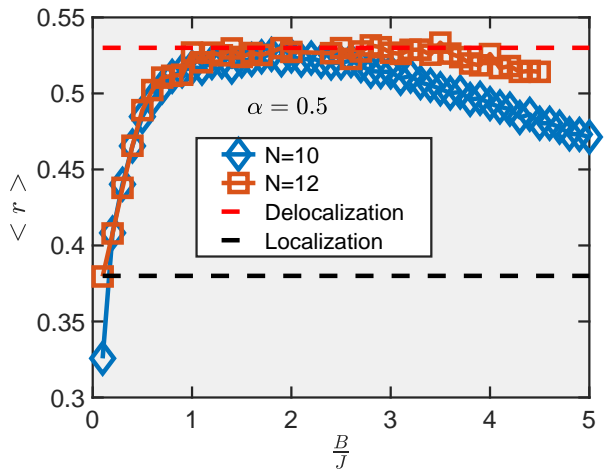}
\caption{\small The level statistics $<r>$ vs. disordering $B$ in transverse fields for $N=10$ and $12$ in the model of Ref. \cite{Hauke15MBLLongRange}.}
\label{fig:RandPot}
\end{figure}

The results change strongly for the middle-energy states considered in Ref. \cite{Li16TransvIs} as described below in Sec. \ref{sec:Middle}. The localization of all states for $\alpha=0.5$ has been also reported in Ref. \cite{Hauke15MBLLongRange} in a slightly modified model compared to Eq. (\ref{eq:H}) (random transverse fields and non-random interactions). 
The preliminary analysis of the level statistics in the model of Ref. \cite{Hauke15MBLLongRange} for the states with nearly zero energy depicted in Fig. \ref{fig:RandPot} indicates the delocalization of these states for $N=12$ for typical transverse fields $B \sim J$. In contrast with Ref. \cite{Hauke15MBLLongRange} we did not use the Kac prescription \cite{Kac63} since it does not affect the appearance of delocalization. 
This result differs from Ref. \cite{Hauke15MBLLongRange} possibly because the contribution of the localized states at energies different from zero included in Ref. \cite{Hauke15MBLLongRange} and  excluded in the present work. The analysis of the problem in detail will be published separately. 

\section{Middle energy states.}
\label{sec:Middle}

We begin the consideration with the analysis of the middle energy states with energies $E_{middle}=(E_{\rm max}+E_{\rm min})/2$ given by the arithmetic average of minimum and maximum eigenstate energies \cite{Li16TransvIs}. The  interaction term in Eq. (\ref{eq:H}) differs from that in the Sherrington-Kirkpatrick spin glass model \cite{Sherrington75}, considered in Ref. \cite{ab16preprintSG}.  In the Sherrington-Kirkpatrick model different interaction constants do not correlate with each other while their correlations in Eq. (\ref{eq:H}) are critically important for the definition of the maximum eigenstate energy. Indeed, both minimum and maximum energies in spin glass model \cite{Sherrington75} are proportional to the number of spins \cite{Aizenman11}, possess opposite signs and approximately equal absolute values  $E_{\rm min}=-E_{\rm max}$. Consequently middle energy states have energies close to zero, which corresponds to infinite temperature. 

\begin{figure}[h!]
\centering
\includegraphics[width=\columnwidth]{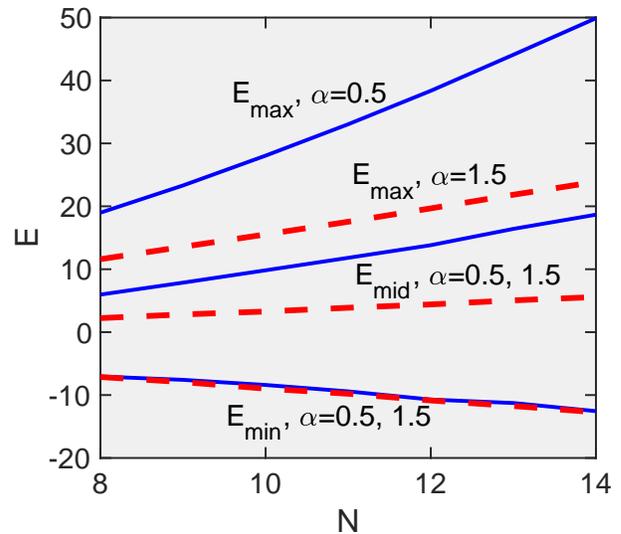}
\caption{\small Dependence of the maximum, minimum and middle energies on the number of spins for $\alpha=0.5$ (straight lines) and $1.5$ (dashed lines) for typical disordering parameter $W=1$.}
\label{fig:Energies}
\end{figure}

The absolute values of the minimum and maximum energies are different for the model given by  Eq. (\ref{eq:H}). The maximum energy corresponds to a sort of ferromagnetic ordering determined as $\sigma_{i}^{z}=1$ (or $-1$) for small disordering $W \ll 1$ or as $\sigma_{i}^{z}={\rm sign} (h_{i})$ (or $-{\rm sign} (h_{i})$) otherwise. For small interaction exponent $\alpha \leq d$ the energy of this state increases with the number of spins superlinearly as ($E_{\rm max} \propto N^{\frac{d-\alpha}{d}}$ for $\alpha < d$ or $E_{\rm max} \propto N\ln(N)$ for $\alpha = d$, remember that $d$ stands for a system dimension). However, the minimum (ground state) energy absolute value increases with the number of spins only proportionally to this number. This is because the antiferromagnetic or spin density wave energy corresponding to the ground state is determined by negative Fourier transforms of spin-spin interaction that don't diverge in the limit of a large number of spins, $N$ \cite{MukamelReviewOneDIsingLongRange}. Consequently, for $\alpha \leq d$ the middle energy increases superlinearly with the number of spins as $E_{middle} \approx E_{\rm max}/2$. This trend is illustrated in Fig. \ref{fig:Energies} for $\alpha=0.5$.

For $\alpha > d$ both minimum and maximum energies are proportional to number of spins yet the maximum energy increases faster (with larger proportionality coefficient) because nearest-neighbor interactions lead to an increase in antiferromagnetic ground state energies.  Consequently, the middle energy increases proportionally to the number of spins as shown in Fig. \ref{fig:Energies} for $\alpha=1.5$. 

\begin{figure}[h!]
\centering
\includegraphics[width=\columnwidth]{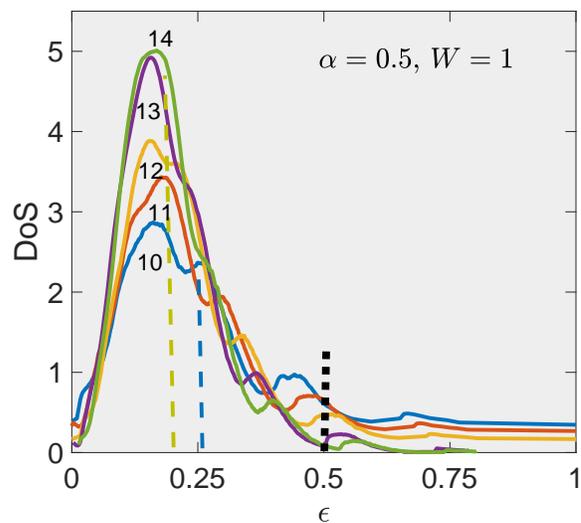}
\caption{\small Density of states (DoS) normalized to $1$ for $\alpha=0.5$ vs. relative energy $\epsilon=(E-E_{\rm min})/(E_{\rm max}-E_{\rm min})$ for typical disordering parameter $W=1$ and different numbers of spins shown near each graph.  Middle energy states with $\epsilon=1/2$ are indicated by dotted line, while two dashed lines restrict the domain corresponding to zero energies at different numbers of particles.}
\label{fig:DoS}
\end{figure}

Consequently, the middle energy states considered in Ref. \cite{Li16TransvIs} correspond to the many-body density of states remarkably smaller than its maximum at energies close to zero. This is illustrated by numerical calculations of the density of states for $\alpha=0.5$ shown in Fig. \ref{fig:DoS}. A remarkable  reduction of the density of states for middle energy states also takes place for other interactions ($\alpha=1$, $1.5$).

\begin{figure}[h!]
\centering
\includegraphics[width=\columnwidth]{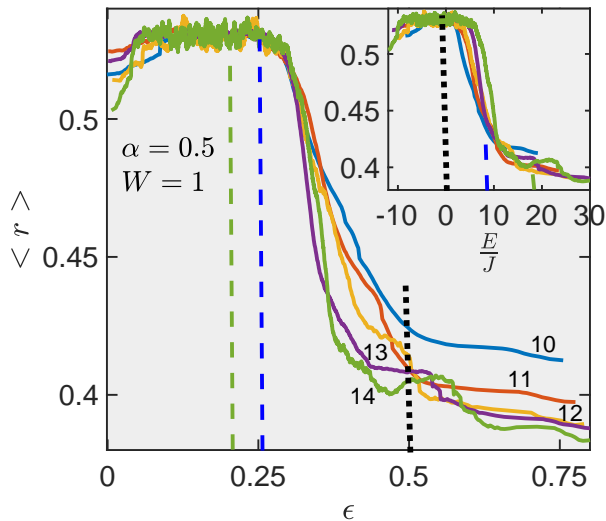}
\caption{\small Level statistics, $<r>$, for $\alpha=0.5$ vs. relative energy $\epsilon=(E-E_{\rm min})/(E_{\rm max}-E_{\rm min})$ and absolute energy (inset) for typical disordering parameter $W=1$ and different numbers of spins shown near each line. The middle energy states and zero energy states are denoted similarly to Fig. \ref{fig:DoS} in the main graph and in the reversed way in the inset.}
\label{fig:OgMiddleW1}
\end{figure}

\begin{figure}[h!]
\centering
\includegraphics[width=\columnwidth]{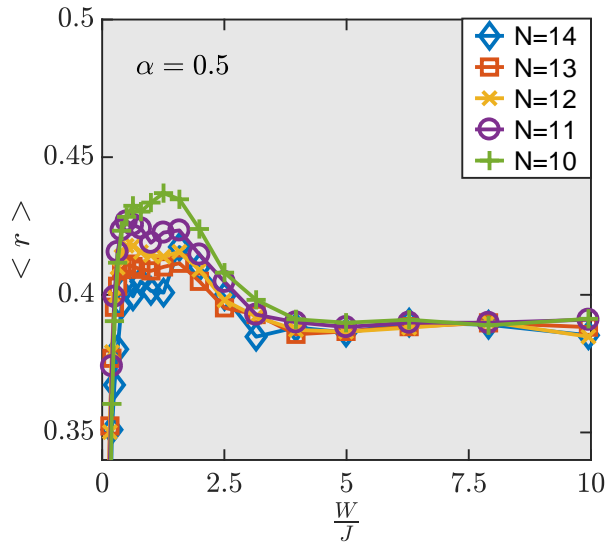}
\caption{\small Level statistics, $<r>$, for $\alpha=0.5$ for middle energy states vs. disordering parameter $W$ for different numbers of spins.}
\label{fig:OgMidvsW}
\end{figure}

It is natural to expect delocalization to be suppressed for states with reduced density \cite{Laumann14,ab16preprintSG}  in agreement with the observations of Ref. \cite{Li16TransvIs}. This is illustrated in Fig. \ref{fig:OgMiddleW1}, where the level statistics parameter, $<r>$, is shown as the function of energy under the same conditions $\alpha=0.5$ and $W=1$ as in  Fig. \ref{fig:DoS}. The plateau at energies close to $0$ corresponding to delocalization ($<r> \approx 0.53$) does not extend to the middle energy states. The average ratio remains significantly smaller than the delocalization limit, $0.53$,  for other disordering parameters $W$ as shown in Fig. \ref{fig:OgMidvsW} reproducing the results of Ref. \cite{Li16TransvIs}. Qualitatively similar behaviors take place for $\alpha=1$.

How does one understand and interpret the delocalization transition for middle energy states having reduced density? Enhanced localization can be naturally expected because of the reduction in the number of accessible states. A very strong enhancement of localization under similar conditions has been found in a random energy model \cite{Laumann14} while in a more realistic spin glass model with binary interactions strong enhancement takes place only below the spin glass phase transition point \cite{ab16preprintSG}, where the majority of spins are substantially frozen out. As discussed below the middle energy states considered in Ref. \cite{Li16TransvIs} indeed correspond to the ordered phase for $\alpha \leq 1$. 

Spin chains with  long-range ($\alpha\leq 2$) ferromagnetic interactions possess finite-temperature phase transition with a broken symmetry resulting in one of two energy minima \cite{Kac1969OneDIsing,Dyson1969OneDIsing,Aizenman1988OneDIsing,
DattaPhTrTransvOneDIsing,Hauke16OneDIsingLongRangeTr} ($\sigma_{i}^{z}=\pm 1$  for  $W \ll 1$ or as $\sigma_{i}^{z}=\pm {\rm sign} (h_{i})$ for $W \gg 1$).

The middle energies most probably correspond to ordered phases in the cases of power-law interactions for $\alpha=0.5$ and $1$. Indeed, the superlinear scaling of middle energies with the number of spins suggests that the majority of spins (around $3/4$) are ferromagnetically ordered for weak disordering, $W \ll 1$, or along random fields $h_{i}$ for strong disordering. Accordingly, the analysis of MBL based on level statistics should be performed with caution since the states with opposite average spin projections are separated by macroscopic barriers.  Consequently, a localization criterion based on level statistics \cite{OganesyanHuse07} may not be applicable. This can explain why the delocalization limit $<r> \approx 0.53$ is not reached for the middle energy states for $\alpha \leq 1$. 

The situation is different in the case of $\alpha=1.5$ where middle energy states possibly correspond to a paramagnetic phase. In this regime the localization threshold should increase with the number of spins as $W_{c} \propto N^{1/4}$ according to Eq. (\ref{eq:LocTrWk}). This expectation conflicts with the statement of Ref. \cite{Li16TransvIs} about an apparent MBL transition point at $W_{c}=6.5$ which is insensitive to the number of spins $N$ (see Fig. 3 there and the related discussion). 

\begin{figure}[h!]
\centering
\includegraphics[width=\columnwidth]{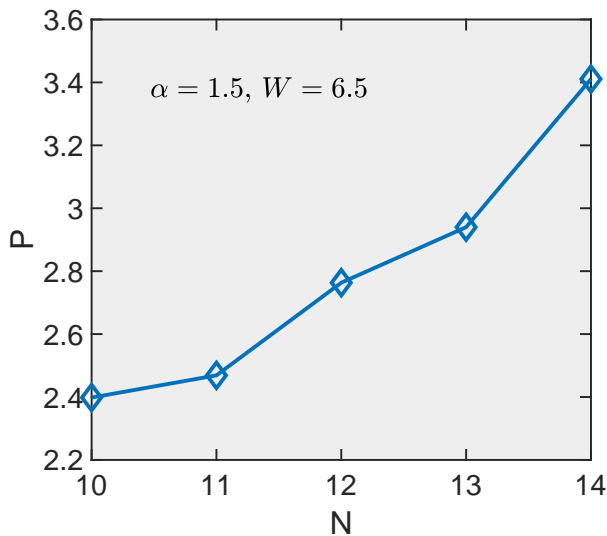}
\caption{\small Participation ratio, $P$, for $\alpha=1.5$ vs the number of spins at disordering parameter $W=6.5$.}
\label{fig:PR1_5Middle}
\end{figure}

To examine the system's behavior at the crossing point we analyzed the eigenstates at related disorder. Each eigenstate of the problem can be represented as a superposition of symmetrized eigenstates of Ising model $|a>$ with coefficients $c_{a}$. The generalized participation ratio $P$ characterizing the number of Ising model states contributed to an individual eigenstate has been evaluated as an inverse geometric average of probabilities $c_{a}^2$ as shown in Fig. \ref{fig:PR1_5Middle}. Using these data one can conclude that the eigenstates are composed of $2$ - $4$ Ising model states out of around $1000$ available states so they are substantially localized in phase (Fock) space. Therefore, in our opinion the estimate $W_{c}=6.5$ of Ref. \cite{Li16TransvIs} is not related to the true delocalization transition.  

This observation raises a general question about the definition of the localization threshold $W_{c}$ using the intersection of dependencies of the average minimum ratios $<r>$ on disordering $W$ at different numbers of spins \cite{Li16TransvIs}. This definition is irrelevant for the present case where the intersection clearly takes place in the localization regime. Therefore other methods \cite{SPAMBC} might be needed in general to determine the localization threshold.

The scaling of the level statistics parameter $<r>$ with the system size at smaller disordering $W$ can possibly be interpreted using its size dependence consistent with Eq. (\ref{eq:LocTrWk}). The predicted dependence $W_{c} \propto N^{1/4}$, Eq. (\ref{eq:LocTrWk}),  is too weak for accurate verification by finite size scaling within the narrow domain of numbers of spins, $10<N<14$, under consideration. This bears investigation 
at larger system sizes.

Partial delocalization of eigenstates in the phase space can take place in the ordered phase for $\alpha \leq 1$ as discussed in Ref. \cite{ab16preprintSG}. The analysis of the participation ratios for middle energies in this case shows a clear signature of delocalization for $\alpha=1$. For instance, the average participation ratios exceed $100$ for weak disorder, $W<2$, in chains of $14$ spins. The deviation of the level statistics from the Wigner Dyson behavior ($<r> < 0.53$ \cite{Li16TransvIs}) in this case is possibly due to the weak coupling of two energy minima $\sigma_{i}=\pm 1$ as discussed above. 

In the case of $\alpha =0.5$ the participation ratio for $10\leq N\leq 14$ and all parameters $W$ does not exceed $10$ strongly indicating localization.  This substantial localization of eigenstates is possibly caused by already large spin flip energies in the ordered phase increasing as $N^{1-\alpha}$ for $\alpha < 1$ ($\sqrt{N}$ for $\alpha=1/2$). This makes delocalization more complicated; yet a preliminary qualitative analysis similar to Ref. \cite{ab15MBL} suggests that all states should be delocalized in the thermodynamic limit for arbitrary disorder $W$. However, the critical disorder increases less strongly with $N$ in the ordered phase (we estimated $W_{c} \propto N^{1/8}$ for $\alpha=0.5$) compared to the paramagnetic phase, Eq. (\ref{eq:LocTrStr}). Therefore numerical observation of this increase requires much longer spin chains. Perhaps this trend can be verified experimentally for spin excitations in cold atoms \cite{Monro16}. 

\section{Conclusion} 

This comment is dedicated to the many-body localization problem in systems with long-range power law interactions $1/r^{\alpha}$, which can be realized experimentally in arrays of cold atoms \cite{Lukin14MBLGen,Lukin14MBLGen,Monro16,LukinDiamond16} with different power law exponents $\alpha$ including $\alpha <d$ in $d$ dimensional systems. It is demonstrated that the systems with $\alpha < d$ inevitably delocalize in the thermodynamic limit, similar to the earlier investigated case $d\leq \alpha < 2d$ and the modified scaling for the critical system size (number of spins) needed for the delocalization is suggested, Eq.  (\ref{eq:LocTrStr}), in the case of infinite temperature. 

The contradiction of this dimensional constraint to the results of the recent paper \cite{Li16TransvIs} for the specific interacting spin model, Eq. (\ref{eq:H}) has been examined. The results conflict with each other for slowly decreasing interaction $\alpha \leq 1$ because the eigenstates of the problem investigated in Ref. \cite{Li16TransvIs}  belong to the ordered phase while the delocalization transition has been misinterpreted for $\alpha =1.5$, i. e. the intersection of the dependencies of level statistic parameters, $<r>$, on disordering, $W$, takes place in the domain of strong localization as follows from the analysis of eigenstate participation ratio. This observation raises a general question about the relevance of this widely-used definition of localization threshold.  

In the ordered phase it is still expected that the inevitable delocalization of eigenstates in the Fock space will take place in thermodynamic limit of large system size at finite temperatures; yet its observation requires much larger sizes than those used in Ref. \cite{Li16TransvIs} which are not hitherto accessible numerically. An experimental verification of the theory  in different settings (e. g. Ref. \cite{Monro16}) is strongly desirable pending numerical investigations 
which may not exceed $N \approx 24$ for the foreseeable future. 



\begin{acknowledgments}
This work is partially supported by the National Science Foundation (CHE-1462075).  
 Authors acknowledge stimulating discussions with Alexander Mirlin and Marcus Heyl. 
\end{acknowledgments}

\bibliography{MBL}

\begin{thebibliography}{27}%
\makeatletter
\providecommand \@ifxundefined [1]{%
 \@ifx{#1\undefined}
}%
\providecommand \@ifnum [1]{%
 \ifnum #1\expandafter \@firstoftwo
 \else \expandafter \@secondoftwo
 \fi
}%
\providecommand \@ifx [1]{%
 \ifx #1\expandafter \@firstoftwo
 \else \expandafter \@secondoftwo
 \fi
}%
\providecommand \natexlab [1]{#1}%
\providecommand \enquote  [1]{``#1''}%
\providecommand \bibnamefont  [1]{#1}%
\providecommand \bibfnamefont [1]{#1}%
\providecommand \citenamefont [1]{#1}%
\providecommand \href@noop [0]{\@secondoftwo}%
\providecommand \href [0]{\begingroup \@sanitize@url \@href}%
\providecommand \@href[1]{\@@startlink{#1}\@@href}%
\providecommand \@@href[1]{\endgroup#1\@@endlink}%
\providecommand \@sanitize@url [0]{\catcode `\\12\catcode `\$12\catcode
  `\&12\catcode `\#12\catcode `\^12\catcode `\_12\catcode `\%12\relax}%
\providecommand \@@startlink[1]{}%
\providecommand \@@endlink[0]{}%
\providecommand \url  [0]{\begingroup\@sanitize@url \@url }%
\providecommand \@url [1]{\endgroup\@href {#1}{\urlprefix }}%
\providecommand \urlprefix  [0]{URL }%
\providecommand \Eprint [0]{\href }%
\providecommand \doibase [0]{http://dx.doi.org/}%
\providecommand \selectlanguage [0]{\@gobble}%
\providecommand \bibinfo  [0]{\@secondoftwo}%
\providecommand \bibfield  [0]{\@secondoftwo}%
\providecommand \translation [1]{[#1]}%
\providecommand \BibitemOpen [0]{}%
\providecommand \bibitemStop [0]{}%
\providecommand \bibitemNoStop [0]{.\EOS\space}%
\providecommand \EOS [0]{\spacefactor3000\relax}%
\providecommand \BibitemShut  [1]{\csname bibitem#1\endcsname}%
\let\auto@bib@innerbib\@empty
\bibitem [{\citenamefont {Burin}(2005)}]{ab06preprint}%
  \BibitemOpen
  \bibfield  {author} {\bibinfo {author} {\bibfnamefont {A.~L.}\ \bibnamefont
  {Burin}},\ }\href@noop {} {\bibfield  {journal} {\bibinfo  {journal}
  {arXiv:cond-mat/0611387}\ } (\bibinfo {year} {2005})}\BibitemShut {NoStop}%
\bibitem [{\citenamefont {Burin}(2015{\natexlab{a}})}]{ab15MBL}%
  \BibitemOpen
  \bibfield  {author} {\bibinfo {author} {\bibfnamefont {A.~L.}\ \bibnamefont
  {Burin}},\ }\href {\doibase 10.1103/PhysRevB.91.094202} {\bibfield  {journal}
  {\bibinfo  {journal} {Phys. Rev. B}\ }\textbf {\bibinfo {volume} {91}},\
  \bibinfo {pages} {094202} (\bibinfo {year} {2015}{\natexlab{a}})}\BibitemShut
  {NoStop}%
\bibitem [{\citenamefont {Li}\ \emph {et~al.}(2016)\citenamefont {Li},
  \citenamefont {Wang}, \citenamefont {Liu},\ and\ \citenamefont
  {Hu}}]{Li16TransvIs}%
  \BibitemOpen
  \bibfield  {author} {\bibinfo {author} {\bibfnamefont {H.}~\bibnamefont
  {Li}}, \bibinfo {author} {\bibfnamefont {J.}~\bibnamefont {Wang}}, \bibinfo
  {author} {\bibfnamefont {X.-J.}\ \bibnamefont {Liu}}, \ and\ \bibinfo
  {author} {\bibfnamefont {H.}~\bibnamefont {Hu}},\ }\href {\doibase
  10.1103/PhysRevA.94.063625} {\bibfield  {journal} {\bibinfo  {journal} {Phys.
  Rev. A}\ }\textbf {\bibinfo {volume} {94}},\ \bibinfo {pages} {063625}
  (\bibinfo {year} {2016})}\BibitemShut {NoStop}%
\bibitem [{\citenamefont {Nandkishore}\ and\ \citenamefont
  {Huse}(2015)}]{Huse15Thermalization}%
  \BibitemOpen
  \bibfield  {author} {\bibinfo {author} {\bibfnamefont {R.}~\bibnamefont
  {Nandkishore}}\ and\ \bibinfo {author} {\bibfnamefont {D.~A.}\ \bibnamefont
  {Huse}},\ }\href {\doibase 10.1146/annurev-conmatphys-031214-014726}
  {\bibfield  {journal} {\bibinfo  {journal} {Annual Review of Condensed Matter
  Physics}\ }\textbf {\bibinfo {volume} {6}},\ \bibinfo {pages} {15} (\bibinfo
  {year} {2015})},\ \Eprint
  {http://arxiv.org/abs/http://dx.doi.org/10.1146/annurev-conmatphys-031214-014726}
  {http://dx.doi.org/10.1146/annurev-conmatphys-031214-014726} \BibitemShut
  {NoStop}%
\bibitem [{\citenamefont {Tikhonenkov}\ \emph {et~al.}(2013)\citenamefont
  {Tikhonenkov}, \citenamefont {Vardi}, \citenamefont {Anglin},\ and\
  \citenamefont {Cohen}}]{Cohen13}%
  \BibitemOpen
  \bibfield  {author} {\bibinfo {author} {\bibfnamefont {I.}~\bibnamefont
  {Tikhonenkov}}, \bibinfo {author} {\bibfnamefont {A.}~\bibnamefont {Vardi}},
  \bibinfo {author} {\bibfnamefont {J.~R.}\ \bibnamefont {Anglin}}, \ and\
  \bibinfo {author} {\bibfnamefont {D.}~\bibnamefont {Cohen}},\ }\href
  {\doibase 10.1103/PhysRevLett.110.050401} {\bibfield  {journal} {\bibinfo
  {journal} {Phys. Rev. Lett.}\ }\textbf {\bibinfo {volume} {110}},\ \bibinfo
  {pages} {050401} (\bibinfo {year} {2013})}\BibitemShut {NoStop}%
\bibitem [{\citenamefont {Huse}\ \emph {et~al.}(2014)\citenamefont {Huse},
  \citenamefont {Nandkishore},\ and\ \citenamefont {Oganesyan}}]{Huse14IntMot}%
  \BibitemOpen
  \bibfield  {author} {\bibinfo {author} {\bibfnamefont {D.~A.}\ \bibnamefont
  {Huse}}, \bibinfo {author} {\bibfnamefont {R.}~\bibnamefont {Nandkishore}}, \
  and\ \bibinfo {author} {\bibfnamefont {V.}~\bibnamefont {Oganesyan}},\ }\href
  {\doibase 10.1103/PhysRevB.90.174202} {\bibfield  {journal} {\bibinfo
  {journal} {Phys. Rev. B}\ }\textbf {\bibinfo {volume} {90}},\ \bibinfo
  {pages} {174202} (\bibinfo {year} {2014})}\BibitemShut {NoStop}%
\bibitem [{\citenamefont {Smith}\ \emph {et~al.}(2015)\citenamefont {Smith},
  \citenamefont {Lee}, \citenamefont {Richerme}, \citenamefont {Neyenhuis},
  \citenamefont {Hess}, \citenamefont {Hauke}, \citenamefont {Heyl},
  \citenamefont {Huse},\ and\ \citenamefont {Monroe}}]{Monro16}%
  \BibitemOpen
  \bibfield  {author} {\bibinfo {author} {\bibfnamefont {J.}~\bibnamefont
  {Smith}}, \bibinfo {author} {\bibfnamefont {A.}~\bibnamefont {Lee}}, \bibinfo
  {author} {\bibfnamefont {P.}~\bibnamefont {Richerme}}, \bibinfo {author}
  {\bibfnamefont {B.}~\bibnamefont {Neyenhuis}}, \bibinfo {author}
  {\bibfnamefont {P.~W.}\ \bibnamefont {Hess}}, \bibinfo {author}
  {\bibfnamefont {P.}~\bibnamefont {Hauke}}, \bibinfo {author} {\bibfnamefont
  {M.}~\bibnamefont {Heyl}}, \bibinfo {author} {\bibfnamefont {D.~A.}\
  \bibnamefont {Huse}}, \ and\ \bibinfo {author} {\bibfnamefont
  {C.}~\bibnamefont {Monroe}},\ }\href@noop {} {\bibfield  {journal} {\bibinfo
  {journal} {arXiv:1508.07026}\ } (\bibinfo {year} {2015})}\BibitemShut
  {NoStop}%
\bibitem [{\citenamefont {Kucsko}\ \emph {et~al.}(2016)\citenamefont {Kucsko},
  \citenamefont {Choi}, \citenamefont {Choi}, \citenamefont {Maurer},
  \citenamefont {Sumiya}, \citenamefont {Onoda}, \citenamefont {Jelezko},
  \citenamefont {Demler}, \citenamefont {Yao},\ and\ \citenamefont
  {Lukin}}]{LukinDiamond16}%
  \BibitemOpen
  \bibfield  {author} {\bibinfo {author} {\bibfnamefont {G.}~\bibnamefont
  {Kucsko}}, \bibinfo {author} {\bibfnamefont {S.}~\bibnamefont {Choi}},
  \bibinfo {author} {\bibfnamefont {J.}~\bibnamefont {Choi}}, \bibinfo {author}
  {\bibfnamefont {P.~C.}\ \bibnamefont {Maurer}}, \bibinfo {author}
  {\bibfnamefont {H.}~\bibnamefont {Sumiya}}, \bibinfo {author} {\bibfnamefont
  {S.}~\bibnamefont {Onoda}}, \bibinfo {author} {\bibfnamefont {J.~I.~F.}\
  \bibnamefont {Jelezko}}, \bibinfo {author} {\bibfnamefont {E.}~\bibnamefont
  {Demler}}, \bibinfo {author} {\bibfnamefont {N.~Y.}\ \bibnamefont {Yao}}, \
  and\ \bibinfo {author} {\bibfnamefont {M.~D.}\ \bibnamefont {Lukin}},\
  }\href@noop {} {\bibfield  {journal} {\bibinfo  {journal} {arXiv:1609.08216}\
  } (\bibinfo {year} {2016})}\BibitemShut {NoStop}%
\bibitem [{\citenamefont {Serbyn}\ \emph {et~al.}(2014)\citenamefont {Serbyn},
  \citenamefont {Knap}, \citenamefont {Gopalakrishnan}, \citenamefont
  {Papi\ifmmode~\acute{c}\else \'{c}\fi{}}, \citenamefont {Yao}, \citenamefont
  {Laumann}, \citenamefont {Abanin}, \citenamefont {Lukin},\ and\ \citenamefont
  {Demler}}]{Lukin14MBLGen}%
  \BibitemOpen
  \bibfield  {author} {\bibinfo {author} {\bibfnamefont {M.}~\bibnamefont
  {Serbyn}}, \bibinfo {author} {\bibfnamefont {M.}~\bibnamefont {Knap}},
  \bibinfo {author} {\bibfnamefont {S.}~\bibnamefont {Gopalakrishnan}},
  \bibinfo {author} {\bibfnamefont {Z.}~\bibnamefont
  {Papi\ifmmode~\acute{c}\else \'{c}\fi{}}}, \bibinfo {author} {\bibfnamefont
  {N.~Y.}\ \bibnamefont {Yao}}, \bibinfo {author} {\bibfnamefont {C.~R.}\
  \bibnamefont {Laumann}}, \bibinfo {author} {\bibfnamefont {D.~A.}\
  \bibnamefont {Abanin}}, \bibinfo {author} {\bibfnamefont {M.~D.}\
  \bibnamefont {Lukin}}, \ and\ \bibinfo {author} {\bibfnamefont {E.~A.}\
  \bibnamefont {Demler}},\ }\href {\doibase 10.1103/PhysRevLett.113.147204}
  {\bibfield  {journal} {\bibinfo  {journal} {Phys. Rev. Lett.}\ }\textbf
  {\bibinfo {volume} {113}},\ \bibinfo {pages} {147204} (\bibinfo {year}
  {2014})}\BibitemShut {NoStop}%
\bibitem [{\citenamefont {Yao}\ \emph {et~al.}(2014)\citenamefont {Yao},
  \citenamefont {Laumann}, \citenamefont {Gopalakrishnan}, \citenamefont
  {Knap}, \citenamefont {M\"uller}, \citenamefont {Demler},\ and\ \citenamefont
  {Lukin}}]{Yao14MBLLongRange}%
  \BibitemOpen
  \bibfield  {author} {\bibinfo {author} {\bibfnamefont {N.~Y.}\ \bibnamefont
  {Yao}}, \bibinfo {author} {\bibfnamefont {C.~R.}\ \bibnamefont {Laumann}},
  \bibinfo {author} {\bibfnamefont {S.}~\bibnamefont {Gopalakrishnan}},
  \bibinfo {author} {\bibfnamefont {M.}~\bibnamefont {Knap}}, \bibinfo {author}
  {\bibfnamefont {M.}~\bibnamefont {M\"uller}}, \bibinfo {author}
  {\bibfnamefont {E.~A.}\ \bibnamefont {Demler}}, \ and\ \bibinfo {author}
  {\bibfnamefont {M.~D.}\ \bibnamefont {Lukin}},\ }\href {\doibase
  10.1103/PhysRevLett.113.243002} {\bibfield  {journal} {\bibinfo  {journal}
  {Phys. Rev. Lett.}\ }\textbf {\bibinfo {volume} {113}},\ \bibinfo {pages}
  {243002} (\bibinfo {year} {2014})}\BibitemShut {NoStop}%
\bibitem [{\citenamefont {Burin}\ \emph {et~al.}(1998)\citenamefont {Burin},
  \citenamefont {Natelson}, \citenamefont {Osheroff},\ and\ \citenamefont
  {Kagan}}]{ab98book}%
  \BibitemOpen
  \bibfield  {author} {\bibinfo {author} {\bibfnamefont {A.~L.}\ \bibnamefont
  {Burin}}, \bibinfo {author} {\bibfnamefont {D.}~\bibnamefont {Natelson}},
  \bibinfo {author} {\bibfnamefont {D.~D.}\ \bibnamefont {Osheroff}}, \ and\
  \bibinfo {author} {\bibfnamefont {Y.}~\bibnamefont {Kagan}},\ }\href@noop {}
  {\emph {\bibinfo {title} {in Tunneling Systems in Amorphous and Crystalline
  Solids}}},\ edited by\ \bibinfo {editor} {\bibfnamefont {P.}~\bibnamefont
  {Esquinazi}}\ (\bibinfo  {publisher} {eds. by P. Esquinazi, p. 223-316
  Springer Verlag},\ \bibinfo {address} {Berlin, Heidelberg, New York},\
  \bibinfo {year} {1998})\ Chap.~\bibinfo {chapter} {5}, pp.\ \bibinfo {pages}
  {223--316}\BibitemShut {NoStop}%
\bibitem [{\citenamefont {Gutman}\ \emph {et~al.}(2016)\citenamefont {Gutman},
  \citenamefont {Protopopov}, \citenamefont {Burin}, \citenamefont {Gornyi},
  \citenamefont {Santos},\ and\ \citenamefont {Mirlin}}]{ab16GutmanMirlin}%
  \BibitemOpen
  \bibfield  {author} {\bibinfo {author} {\bibfnamefont {D.~B.}\ \bibnamefont
  {Gutman}}, \bibinfo {author} {\bibfnamefont {I.~V.}\ \bibnamefont
  {Protopopov}}, \bibinfo {author} {\bibfnamefont {A.~L.}\ \bibnamefont
  {Burin}}, \bibinfo {author} {\bibfnamefont {I.~V.}\ \bibnamefont {Gornyi}},
  \bibinfo {author} {\bibfnamefont {R.~A.}\ \bibnamefont {Santos}}, \ and\
  \bibinfo {author} {\bibfnamefont {A.~D.}\ \bibnamefont {Mirlin}},\ }\href
  {\doibase 10.1103/PhysRevB.93.245427} {\bibfield  {journal} {\bibinfo
  {journal} {Phys. Rev. B}\ }\textbf {\bibinfo {volume} {93}},\ \bibinfo
  {pages} {245427} (\bibinfo {year} {2016})}\BibitemShut {NoStop}%
\bibitem [{\citenamefont {Burin}(2015{\natexlab{b}})}]{ab15MBLXY}%
  \BibitemOpen
  \bibfield  {author} {\bibinfo {author} {\bibfnamefont {A.~L.}\ \bibnamefont
  {Burin}},\ }\href {\doibase 10.1103/PhysRevB.92.104428} {\bibfield  {journal}
  {\bibinfo  {journal} {Phys. Rev. B}\ }\textbf {\bibinfo {volume} {92}},\
  \bibinfo {pages} {104428} (\bibinfo {year} {2015}{\natexlab{b}})}\BibitemShut
  {NoStop}%
\bibitem [{\citenamefont {Oganesyan}\ and\ \citenamefont
  {Huse}(2007)}]{OganesyanHuse07}%
  \BibitemOpen
  \bibfield  {author} {\bibinfo {author} {\bibfnamefont {V.}~\bibnamefont
  {Oganesyan}}\ and\ \bibinfo {author} {\bibfnamefont {D.~A.}\ \bibnamefont
  {Huse}},\ }\href {\doibase 10.1103/PhysRevB.75.155111} {\bibfield  {journal}
  {\bibinfo  {journal} {Phys. Rev. B}\ }\textbf {\bibinfo {volume} {75}},\
  \bibinfo {pages} {155111} (\bibinfo {year} {2007})}\BibitemShut {NoStop}%
\bibitem [{\citenamefont {Hauke}\ and\ \citenamefont
  {Heyl}(2015)}]{Hauke15MBLLongRange}%
  \BibitemOpen
  \bibfield  {author} {\bibinfo {author} {\bibfnamefont {P.}~\bibnamefont
  {Hauke}}\ and\ \bibinfo {author} {\bibfnamefont {M.}~\bibnamefont {Heyl}},\
  }\href {\doibase 10.1103/PhysRevB.92.134204} {\bibfield  {journal} {\bibinfo
  {journal} {Phys. Rev. B}\ }\textbf {\bibinfo {volume} {92}},\ \bibinfo
  {pages} {134204} (\bibinfo {year} {2015})}\BibitemShut {NoStop}%
\bibitem [{\citenamefont {Laumann}\ \emph {et~al.}(2014)\citenamefont
  {Laumann}, \citenamefont {Pal},\ and\ \citenamefont
  {Scardicchio}}]{Laumann14}%
  \BibitemOpen
  \bibfield  {author} {\bibinfo {author} {\bibfnamefont {C.~R.}\ \bibnamefont
  {Laumann}}, \bibinfo {author} {\bibfnamefont {A.}~\bibnamefont {Pal}}, \ and\
  \bibinfo {author} {\bibfnamefont {A.}~\bibnamefont {Scardicchio}},\ }\href
  {\doibase 10.1103/PhysRevLett.113.200405} {\bibfield  {journal} {\bibinfo
  {journal} {Phys. Rev. Lett.}\ }\textbf {\bibinfo {volume} {113}},\ \bibinfo
  {pages} {200405} (\bibinfo {year} {2014})}\BibitemShut {NoStop}%
\bibitem [{\citenamefont {Burin}(2017)}]{ab16preprintSG}%
  \BibitemOpen
  \bibfield  {author} {\bibinfo {author} {\bibfnamefont {A.}~\bibnamefont
  {Burin}},\ }\href {\doibase 10.1002/andp.201600292} {\bibfield  {journal}
  {\bibinfo  {journal} {Annalen der Physik}\ ,\ \bibinfo {pages} {1600292}}
  (\bibinfo {year} {2017})},\ \bibinfo {note} {1600292}\BibitemShut {NoStop}%
\bibitem [{\citenamefont {Kac}\ \emph {et~al.}(1963)\citenamefont {Kac},
  \citenamefont {Uhlenbeck},\ and\ \citenamefont {Hemmer}}]{Kac63}%
  \BibitemOpen
  \bibfield  {author} {\bibinfo {author} {\bibfnamefont {M.}~\bibnamefont
  {Kac}}, \bibinfo {author} {\bibfnamefont {G.~E.}\ \bibnamefont {Uhlenbeck}},
  \ and\ \bibinfo {author} {\bibfnamefont {P.~C.}\ \bibnamefont {Hemmer}},\
  }\href {\doibase http://dx.doi.org/10.1063/1.1703946} {\bibfield  {journal}
  {\bibinfo  {journal} {Journal of Mathematical Physics}\ }\textbf {\bibinfo
  {volume} {4}},\ \bibinfo {pages} {216} (\bibinfo {year} {1963})}\BibitemShut
  {NoStop}%
\bibitem [{\citenamefont {Sherrington}\ and\ \citenamefont
  {Kirkpatrick}(1975)}]{Sherrington75}%
  \BibitemOpen
  \bibfield  {author} {\bibinfo {author} {\bibfnamefont {D.}~\bibnamefont
  {Sherrington}}\ and\ \bibinfo {author} {\bibfnamefont {S.}~\bibnamefont
  {Kirkpatrick}},\ }\href {\doibase 10.1103/PhysRevLett.35.1792} {\bibfield
  {journal} {\bibinfo  {journal} {Phys. Rev. Lett.}\ }\textbf {\bibinfo
  {volume} {35}},\ \bibinfo {pages} {1792} (\bibinfo {year}
  {1975})}\BibitemShut {NoStop}%
\bibitem [{\citenamefont {Aizenman}\ and\ \citenamefont
  {Warzel}(2011)}]{Aizenman11}%
  \BibitemOpen
  \bibfield  {author} {\bibinfo {author} {\bibfnamefont {M.}~\bibnamefont
  {Aizenman}}\ and\ \bibinfo {author} {\bibfnamefont {S.}~\bibnamefont
  {Warzel}},\ }\href {\doibase 10.1103/PhysRevLett.106.136804} {\bibfield
  {journal} {\bibinfo  {journal} {Phys. Rev. Lett.}\ }\textbf {\bibinfo
  {volume} {106}},\ \bibinfo {pages} {136804} (\bibinfo {year}
  {2011})}\BibitemShut {NoStop}%
\bibitem [{\citenamefont {Mukamel}(2008)}]{MukamelReviewOneDIsingLongRange}%
  \BibitemOpen
  \bibfield  {author} {\bibinfo {author} {\bibfnamefont {D.}~\bibnamefont
  {Mukamel}},\ }\href {\doibase 10.1063/1.2839123} {\bibfield  {journal}
  {\bibinfo  {journal} {AIP Conference Proceedings}\ }\textbf {\bibinfo
  {volume} {970}},\ \bibinfo {pages} {22} (\bibinfo {year} {2008})},\ \Eprint
  {http://arxiv.org/abs/http://aip.scitation.org/doi/pdf/10.1063/1.2839123}
  {http://aip.scitation.org/doi/pdf/10.1063/1.2839123} \BibitemShut {NoStop}%
\bibitem [{\citenamefont {Kac}\ and\ \citenamefont
  {Thompson}(1969)}]{Kac1969OneDIsing}%
  \BibitemOpen
  \bibfield  {author} {\bibinfo {author} {\bibfnamefont {M.}~\bibnamefont
  {Kac}}\ and\ \bibinfo {author} {\bibfnamefont {C.~J.}\ \bibnamefont
  {Thompson}},\ }\href {\doibase 10.1063/1.1664976} {\bibfield  {journal}
  {\bibinfo  {journal} {Journal of Mathematical Physics}\ }\textbf {\bibinfo
  {volume} {10}},\ \bibinfo {pages} {1373} (\bibinfo {year} {1969})},\ \Eprint
  {http://arxiv.org/abs/http://dx.doi.org/10.1063/1.1664976}
  {http://dx.doi.org/10.1063/1.1664976} \BibitemShut {NoStop}%
\bibitem [{\citenamefont {Dyson}(1969)}]{Dyson1969OneDIsing}%
  \BibitemOpen
  \bibfield  {author} {\bibinfo {author} {\bibfnamefont {F.~J.}\ \bibnamefont
  {Dyson}},\ }\href {http://projecteuclid.org/euclid.cmp/1103841344} {\bibfield
   {journal} {\bibinfo  {journal} {Comm. Math. Phys.}\ }\textbf {\bibinfo
  {volume} {12}},\ \bibinfo {pages} {91} (\bibinfo {year} {1969})}\BibitemShut
  {NoStop}%
\bibitem [{\citenamefont {Aizenman}\ \emph {et~al.}(1988)\citenamefont
  {Aizenman}, \citenamefont {Chayes}, \citenamefont {Chayes},\ and\
  \citenamefont {Newman}}]{Aizenman1988OneDIsing}%
  \BibitemOpen
  \bibfield  {author} {\bibinfo {author} {\bibfnamefont {M.}~\bibnamefont
  {Aizenman}}, \bibinfo {author} {\bibfnamefont {J.~T.}\ \bibnamefont
  {Chayes}}, \bibinfo {author} {\bibfnamefont {L.}~\bibnamefont {Chayes}}, \
  and\ \bibinfo {author} {\bibfnamefont {C.~M.}\ \bibnamefont {Newman}},\
  }\href {\doibase 10.1007/BF01022985} {\bibfield  {journal} {\bibinfo
  {journal} {Journal of Statistical Physics}\ }\textbf {\bibinfo {volume}
  {50}},\ \bibinfo {pages} {1} (\bibinfo {year} {1988})}\BibitemShut {NoStop}%
\bibitem [{\citenamefont {Dutta}\ and\ \citenamefont
  {Bhattacharjee}(2001)}]{DattaPhTrTransvOneDIsing}%
  \BibitemOpen
  \bibfield  {author} {\bibinfo {author} {\bibfnamefont {A.}~\bibnamefont
  {Dutta}}\ and\ \bibinfo {author} {\bibfnamefont {J.~K.}\ \bibnamefont
  {Bhattacharjee}},\ }\href {\doibase 10.1103/PhysRevB.64.184106} {\bibfield
  {journal} {\bibinfo  {journal} {Phys. Rev. B}\ }\textbf {\bibinfo {volume}
  {64}},\ \bibinfo {pages} {184106} (\bibinfo {year} {2001})}\BibitemShut
  {NoStop}%
\bibitem [{\citenamefont {Zunkovic}\ \emph {et~al.}(2016)\citenamefont
  {Zunkovic}, \citenamefont {Heyl}, \citenamefont {Knap},\ and\ \citenamefont
  {Silva}}]{Hauke16OneDIsingLongRangeTr}%
  \BibitemOpen
  \bibfield  {author} {\bibinfo {author} {\bibfnamefont {B.}~\bibnamefont
  {Zunkovic}}, \bibinfo {author} {\bibfnamefont {M.}~\bibnamefont {Heyl}},
  \bibinfo {author} {\bibfnamefont {M.}~\bibnamefont {Knap}}, \ and\ \bibinfo
  {author} {\bibfnamefont {A.}~\bibnamefont {Silva}},\ }\href@noop {}
  {\bibfield  {journal} {\bibinfo  {journal} {arXiv:1609.08482
  [cond-mat.quant-gas]}\ } (\bibinfo {year} {2016})}\BibitemShut {NoStop}%
\bibitem [{\citenamefont {Serbyn}\ \emph {et~al.}(2015)\citenamefont {Serbyn},
  \citenamefont {Papi\ifmmode~\acute{c}\else \'{c}\fi{}},\ and\ \citenamefont
  {Abanin}}]{SPAMBC}%
  \BibitemOpen
  \bibfield  {author} {\bibinfo {author} {\bibfnamefont {M.}~\bibnamefont
  {Serbyn}}, \bibinfo {author} {\bibfnamefont {Z.}~\bibnamefont
  {Papi\ifmmode~\acute{c}\else \'{c}\fi{}}}, \ and\ \bibinfo {author}
  {\bibfnamefont {D.~A.}\ \bibnamefont {Abanin}},\ }\href {\doibase
  10.1103/PhysRevX.5.041047} {\bibfield  {journal} {\bibinfo  {journal} {Phys.
  Rev. X}\ }\textbf {\bibinfo {volume} {5}},\ \bibinfo {pages} {041047}
  (\bibinfo {year} {2015})}\BibitemShut {NoStop}%
\end{thebibliography}%

\end{document}